\documentclass[floatfix,amssymb,longbibliography,amsfonts,amsmath,nopacs,twocolumn,pre,aps]{revtex4-1}

\def\TReg{\textsuperscript{\textregistered}}

\usepackage{helvet}
\usepackage{courier}

\usepackage[version=3]{mhchem} 

\usepackage{german,latexsym}
\usepackage{amsmath, amssymb, wasysym}
\usepackage{bbm}
\usepackage{color}
\usepackage{rotating}
\usepackage{tabularx}
\newcolumntype{C}[1]{>{\centering\arraybackslash}p{#1}}

\usepackage{longtable}		
\usepackage{multirow}

\definecolor{orange}{rgb}{1,0.65,0}

\usepackage{pifont}
\usepackage{ifsym}
\usepackage{textcomp}
\usepackage{units} 

\usepackage[latin1]{inputenc}
\usepackage[english]{babel}

\usepackage[pdftex,
						colorlinks=true,
						breaklinks=true,
						linkcolor=black,
				    urlcolor=black,
						citecolor=black,
						hyperindex,
						pdfdisplaydoctitle=true,
						pdfpagemode=UseOutlines,
						pdfkeywords={dynamics, rheology, water, alkanes, polymer, liquid crystals, imbibition, neutron scattering},
						pdftitle={Rheology and Dynamics of Simple and Complex Liquids in nanoporous Matrices},
						pdfauthor={Simon A. Gruener},
						pdfsubject={Dissertation},
						plainpages=false,						
						pdfpagelabels=true]
						{hyperref}
\usepackage{hypernat}

\usepackage{lmodern} 
\usepackage[T1]{fontenc} 

\newcommand{\degr}{$^{\circ}$}
\newcommand{\dC}{\,$^{\circ}$C}

\newcommand{\subs}[2]{#1_{\mbox{{\scriptsize #2}}}}
\newcommand{\V}{Vy\-cor\TReg\ }
\newcommand{\Vend}{Vy\-cor\TReg}

\newcommand{\etal}{\textit{et al.} }

\newcommand{\bild}[1]{Fig.~\ref{#1}}
\newcommand{\tab}[1]{Tab.~\ref{#1}}
\newcommand{\rel}[1]{Eq.~\eqref{#1}}

\newcommand{\etalp}{\textit{et~al.~}}

\selectfont

\begin{document}
\title{Hydraulic Transport Across Hydrophilic and Hydrophobic Nanopores:\\ Flow Experiments with Water and n-Hexane}
\author{Simon Gruener}
\email[E-mail: ]{simon-alexander.gruener@basf.com}
\affiliation{Experimental Physics, Saarland University, D-66041 Saarbrücken (Germany)}
\affiliation{Sorption and Permeation Laboratory, BASF SE, D-67056 Ludwigshafen (Germany)}
\author{Dirk Wallacher}
\affiliation{Experimental Physics, Saarland University, D-66041 Saarbrücken (Germany)}
\affiliation{Department Sample Environments, Helmholtz-Centre Berlin for Energy and Materials, Hahn-Meitner-Platz 1, D-14109 Berlin (Germany)}
\author{Stefanie Greulich}
\affiliation{Experimental Physics, Saarland University, D-66041 Saarbrücken (Germany)}
\author{Mark Busch}
\affiliation{Institute of Materials Physics and Technology, Ei\ss endorfer Str. 42, D-21073 Hamburg-Harburg (Germany)}

\author{Patrick Huber}
\email[E-mail: ]{patrick.huber@tuhh.de}
\affiliation{Experimental Physics, Saarland University, D-66041 Saarbrücken (Germany)}
\affiliation{Institute of Materials Physics and Technology, Ei\ss endorfer Str. 42, D-21073 Hamburg-Harburg (Germany)}

\date{\today}

\begin{abstract}
We experimentally explore pressure-driven flow of water and n-hexane across nanoporous silica (\V glass monoliths with 7 or 10~nm pore diameters, respectively) as a function of temperature and surface functionalization (native and silanized glass surfaces). Hydraulic flow rates are measured by applying hydrostatic pressures via inert gases (argon and helium, pressurized up to 70 bar) on the upstream side in a capacitor-based membrane permeability setup. For the native, hydrophilic silica walls, the measured hydraulic permeabilities can be quantitatively accounted for by bulk fluidity provided we assume a sticking boundary layer, i.e. a negative velocity slip length of molecular dimensions. The thickness of this boundary layer is discussed with regard to previous capillarity-driven flow experiments (spontaneous imbibition) and with regard to velocity slippage at the pore walls resulting from dissolved gas. Water flow across the silanized, hydrophobic nanopores is blocked up to a hydrostatic pressure of at least 70 bar. The absence of a sticking boundary layer quantitatively accounts for an enhanced n-hexane permeability in the hydrophobic compared to the hydrophilic nanopores.  
\end{abstract}


\maketitle

\section{Introduction}

Liquid flow and shear in pores a few nanometers across plays a dominant role in a plethora of processes and phenomena encompassing transport across biomembranes and biological tissues \cite{Eijkel2005, Stroock2014, Mastrangeli2015}, geological erosion and hydraulic fracturing \cite{Birdsell2015, Falk2015}, the synthesis of nanostructured hybrid materials by electrodeposition \cite{Yin2001} or melt infiltration \cite{Huczko2000, Yin2001, Steinhart2002, Sander2003, Jongh2013, Huber2015}, the separation of liquids by filter membranes, the durability of concrete \cite{Pleinert1998, Vichit-Vadakan2002, Zhang2011} and friction \cite{Rosenhek-Goldian2015,Jee2015,Israelachvili2015}. The possibility for energy storage by forced liquid intrusion in nanoporous media is another topic which increasingly attracts interest both from a fundumental and an applied perspective \cite{Michelin-Jamois2015, Lefevre2004, Grosu2014, Borman2015}. 

Also the goal for an engineering of flows of minute amount of liquids in small devices, i.e. the design of lab-on-a-chip devices \cite{Stone2004, Eijkel2005, Squires2005, Majumder2005, Dittrich2006, Whitby2007, Karnik2006, Persson2007, Schoch2008,  Muller2008, Piruska2010, Kirby2010, Bocquet2010, Koester2012, Duan2013, Bocquet2014, Vincent2014, Xue2014, Tani2015, Vincent2015, Li2015} motivates research activities with regard to the flow properties of liquids in extreme spatial confinement.

Similarly as for the thermodynamic equilibrium properties of pore-confined condensed matter 
 \cite{Huber1999, Gelb1999, Christenson2001, Alba-Simionesco2006, Knorr2008, Huber2015} a couple of interesting questions regarding the transport behavior for the flow in such restricted geometries arise \cite{Gruener2011}: (i) Can the macroscopic wetting properties or values of fluid parameters, such as the viscosity $\eta$, surface and interfacial tensions $\sigma$, accurately describe a liquid down to very small length scales, on the order of the size of its building blocks \cite{Fradin2000, Vinogradova2011, Baeumchen2014, Bocquet2014}. (ii) What happens with the conventional hydrodynamic no-slip shear stress boundary condition at the confining walls? (iii) How sensitive depends the nanofluidic transport behavior on dissolved gases?

Measurements with the surface-force apparatus (SFA), which allow one to study shear viscosities \cite{Chan1985, Christenson1982, Stevens1997, Georges1993, Heinbuch1989, Ruths2000, Raviv2001} and frictional properties \cite{Rosenhek-Goldian2015,Jee2015,Israelachvili2015} of thin films with thicknesses down to sub-nanometers, have revealed that depending on the shear rate, the type of molecule and the surface chemistry sometimes remarkably robust bulk fluidity could be observed, sometimes however also sizeable deviations. 

In general, the enormous academic and economic interests on the interfacial behavior of liquids are manifested by a vast publication rate concerning this issue during the last decade. Many different techniques like SFA, atomic force microscopy, particle image velocimetry, fluorescence recovery after photobleaching and controlled dewetting as well as molecular dynamics or lattice Boltzmann simulations were utilized \cite{Neto2005, Harting2010, Baeumchen2010,Chen2012, Sedghi2014,Bocquet2014}. 

Up to date many factors have been found that seem to influence the boundary conditions. The most prominent and maybe the least controversially discussed amongst them is the fluid-wall interaction expressed in terms of the wettability \cite{Barrat99, Pit00, Cieplak01, Tretheway02, Cho04, Schmatko05, Fetzer2007, Voronov2008, Maali2008, Servantie2008, Muller2008, Cao2009, Baeumchen2010, Ho2011,Schaeffel2013, Bhadauria2013, Sedghi2014}. The weaker the interaction is the more likely is slip. In addition, shear rates beyond a critical value are supposed to induce slip, too \cite{Thompson1997, Zhu01, Craig01, Priezjev04, Priezjev2007}. In contrast, the influence of surface roughness is rather debatable \cite{Vinogradova06}. There are results for a decrease \cite{Zhu2002, Pit2000} as well as for an increase \cite{Bonaccurso2003} of the slip length with increasing surface roughness. Furthermore, dissolved gases \cite{Granick2003, Dammer2006, Harting2010, Yen2014}, the shape of the fluid molecules \cite{Schmatko2005} or the add-on of surfactants \cite{Cheikh2003} seem to influence the boundary conditions. To sum up, there is a huge set of factors (see Refs.~\cite{Granick2003, Lauga2005, Neto2005, Bocquet2014}) and certainly a complex interplay between many of them finally determines the interfacial flow behavior. 

Pioneering experiments to probe transport behavior through nanoporous media were performed by Nordberg \cite{Nordberg1944} and Debye and Cleland in the mid of the last century \cite{Debye1959}. Nordberg studied water and acetone flow, whereas Debye and Cleland reported on the flow of a series of linear hydrocarbons (n-pentane to n-octadecane) through nanoporous \V glass. Flow rates in agreement with Darcy's law, the generalisation of Hagen-Poiseuille's law for simple capillaries towards porous media \cite{Gruener2009}, were observed. 

As Abeles \etalp \cite{Abeles1991} documented by an experimental study on toluene using also nanoporous \V glass, flow in nanoporous media can be through molecular flow (also termed Knudsen diffusion \cite{Gruener2008}), surface diffusion, and viscous liquid flow driven by capillary forces (termed ''spontaneous imbibition'')\cite{Gruener2009, Gruener2012, Gruener2015} or by external hydraulic pressure (called ''forced imbibition'') \cite{Alava2004, Dimitrov2008a}, depending on the size of the pores and on the temperature and pressure of the fluid. 

In the following we focus on forced imbibition dynamics of water and n-hexane across monolithic \V glass monoliths with two distinct mean pore diameters. After a short introduction to flow across sponge-like porous media, we present our experimental setup and discuss our results with regard to the hydrodynamic boundary conditions at the pore walls as a function of surface functionalization and the possible influence of dissolved gases.

\section[Principles of liquid flows]{{Principles of liquid flows in a nanoporous medium}}
The measurements presented below involve the flow of liquids through \V, a complex pore network comparable to a sponge. As simple approximation one can reduce the problem to the flow of a fluid through a tiny capillary. Consequently, the law of Hagen-Poiseuille is the starting point of the subsequent development of a theory of the liquid flow in a pore network.

\subsection[Liquid flow in isotropic networks]{Liquid Flow in Isotropic Pore Networks}
For a given pressure difference $\Delta p$\label{pressuredifference} applied along a cylindrical duct with radius $r$\label{radius1} and length $\ell$ the volume flow rate $\dot{V}$ is determined by
\begin{equation}
\dot{V} = \frac{\pi \,r^4 }{8\,\eta \,\ell} \,\Delta p \; .
\label{eq:HagenPoiseuille}
\end{equation}
Here $\eta$\label{viscosity} denotes the dynamic viscosity of the flowing liquid. In the next step, one has to evolve concepts in order to account for the sponge-like structure of an isotropic pore network. In general such a network can be characterized by three quantities. The mean pore radius $\subs{r}{0}$\label{meanporeradius} and the volume porosity $\subs{\phi}{0}$\label{volumeporosity}, as obtained from sorption isotherm experiments, are probably the most intuitive ones among them. With only these two parameters a porous cuboid with edge length $a$ (and cross-sectional area $A=a^2$) consisting of\label{crosssectionalarea}
\begin{equation}
n=\frac{\subs{\phi}{0}\,A}{\pi\,\subs{r}{0}^2} \qquad \left(\;\Leftrightarrow \;\; \subs{\phi}{0} \equiv \frac{\subs{V}{void}}{\subs{V}{sample}} = \frac{n\,\pi\,\subs{r}{0}^2\,a}{a^3} \; \right)
\label{eq:porenumber}
\end{equation}
cylindrical pores with radius $\subs{r}{0}$ and length $a$ can be constructed. Assuming the capillaries to be aligned in flow direction the flow rate through the whole matrix is then given by $n$ times the single pore flow rate Eq.~(\ref{eq:HagenPoiseuille}) with $r=\subs{r}{0}$ and $\ell=a$. However, so far this description still lacks information on the orientation of the pores. 

\begin{figure}[!b]
\centering
\includegraphics*[width=.4\linewidth]{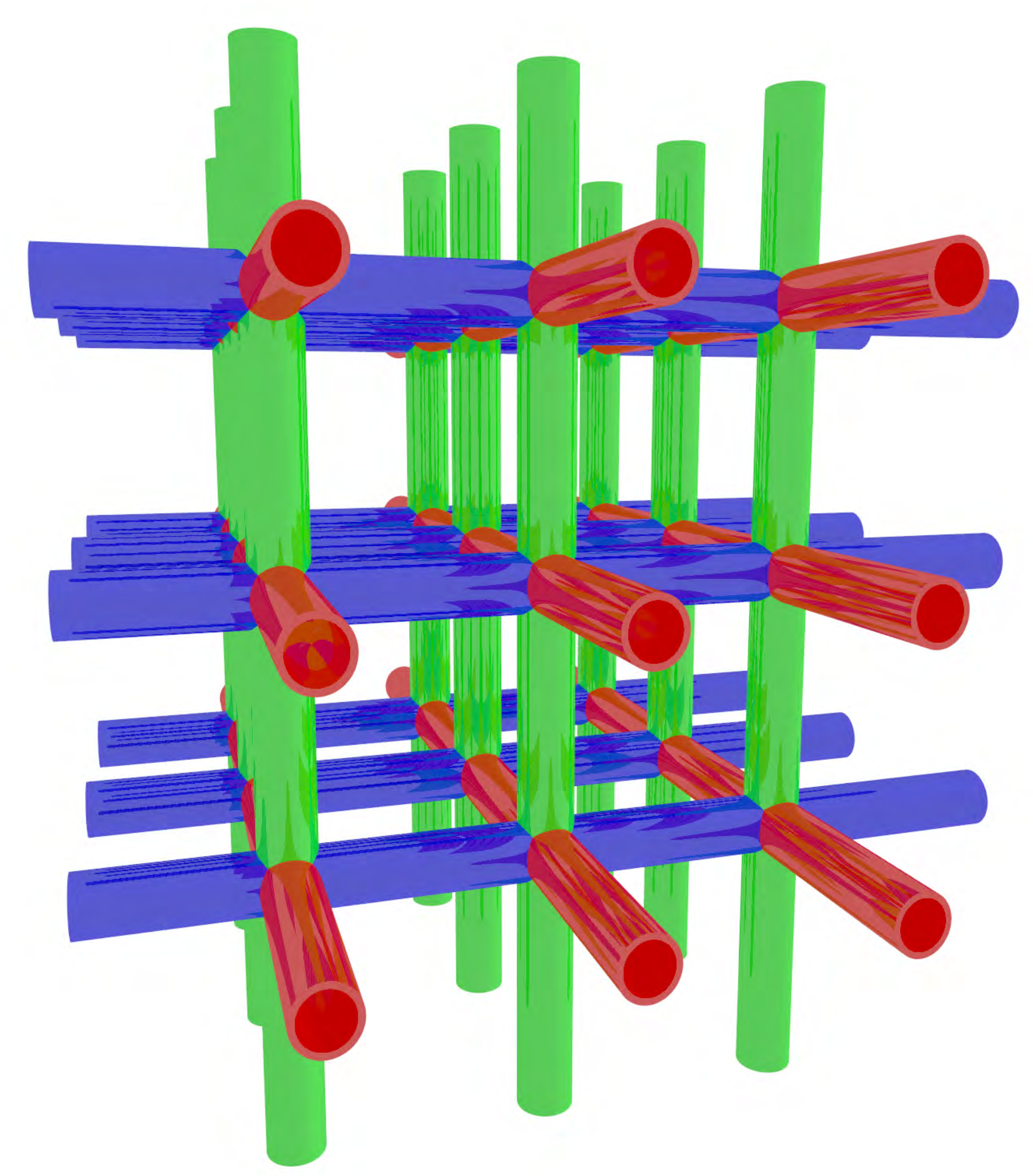}
\hspace{.1\linewidth}
\includegraphics*[width=.45\linewidth]{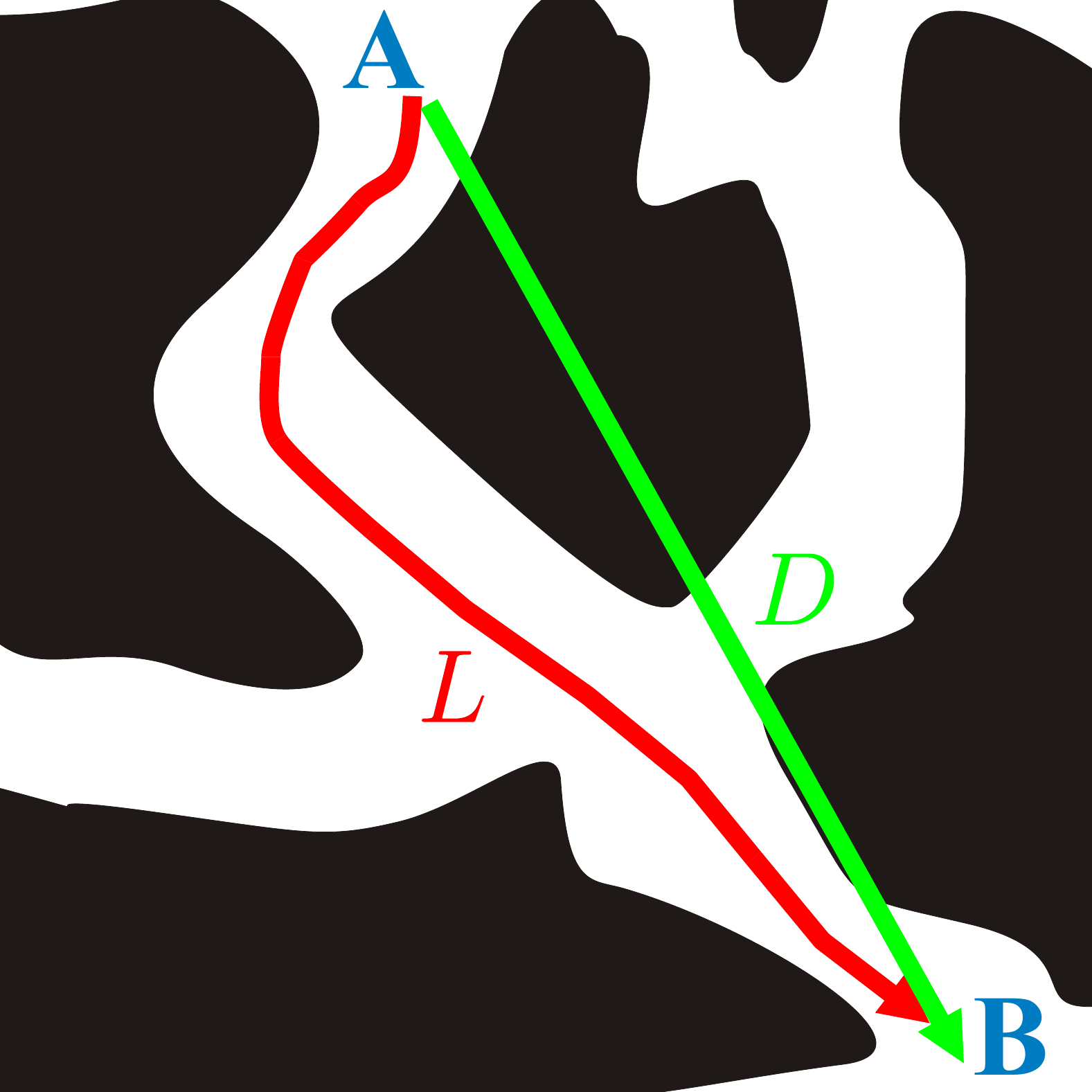}
\caption[Illustration of the meaning of the tortuosity of a pore network]{(Color online) Illustration of the meaning of the tortuosity $\tau$ of a pore network such as \Vend. (left): For an isotropic distribution of the pores only every third pore is subjected to the pressure gradient yielding $\tau=3$. (right): For meandering pores an additional factor of $\frac{L}{D}$ must be introduced to correct the length $D$ of the direct interconnection of two points for the actual path length $L$.}
\label{pic_tortuosity}
\end{figure}

To account for the isotropy of the network as indicated in Fig.~\ref{pic_tortuosity}\,(left) it is necessary to introduce a third parameter, the so-called tortuosity $\tau$\label{tortuosity} along with the transformation
\begin{equation}
\dot{V} \quad \longrightarrow \quad \frac{1}{\tau} \dot{V}
\label{eq:tautransform}
\end{equation}
of the volume flow rate Eq.~(\ref{eq:HagenPoiseuille}). Pores totally aligned in flow direction would yield $\tau=1$, whereas isotropic distributed pores would result in $\tau=3$. For a random orientation only every third pore is subjected to the pressure gradient and hence contributes to the flow. Therefore, the net flow rate has to be divided by the factor three. But no correction is needed if all pores are aligned in flow direction and as a result of this it is $\tau=1$. In this way the tortuosity is a simple method for accounting for the orientation of the pores with respect to the direction of the pressure drop.

To date several techniques have been applied to extract the tortuosity of the isotropic pore network in \V glass. Deducing the diffusion coefficient of hexane and decane by means of small angle neutron scattering (SANS) measurements $\tau$ was found to be in the range of 3.4 - 4.2 \cite{Lin92}. Gas permeation measurements performed with an in-house apparatus resulted in $\tau = 3.9\pm 0.4$ \cite{Gruener2008, Bommer08}. Finally, calculations based on three-dimensional geometrical models yielded a value of approximately 3.5 \cite{Crossley91}. 

Interestingly, all values show a significant deviation from $\tau=3$ as derived from the previous considerations. Accordingly, there must be an additional aspect of the geometry that has so far been neglected. Regarding Fig.~\ref{pic_tortuosity}\,(right) this issue is apparent: the pores are not straight but rather meandering. In consequence the length $L$ of the path from any point A to another point B is always larger than the length $D$ of the direct interconnection of the two points. To correct the pore length for the larger flow path an additional factor of $\frac{L}{D}$ for the tortuosity must be introduced. Assuming $\tau =3.6$ this consideration yields for the \V pore network $L\approx 1.2\, D$. This result can vividly be interpreted as follows: the shortest way from the bottom of the previously introduced sample cuboid to its top is about 20\,\% longer than its edge length $a$.
With all the preceding considerations in mind one is able to derive an expression that describes the flow of a liquid through a porous network. For a given porous matrix with cross-sectional area $A$ and thickness $d$\label{samplethickness} (along which the pressure drop $\Delta p$ is applied) the normalized volume flow rate $\frac{1}{A}\dot{V}$ is determined by
\begin{equation}
\frac{1}{A} \, \dot{V} = \frac{K}{\eta\,d} \,\Delta p \; .
\label{eq:Darcy}
\end{equation}
This expression is also known as Darcy's law \cite{Debye1959}. The proportionality constant $K$ is the so-called hydraulic permeability\label{permeability} of the matrix. It is given by
\begin{equation}
K=\frac{\subs{\phi}{0}}{8\, \tau} \,\subs{r}{0}^2 \, .
\label{eq:permeability}
\end{equation}
Note that the permeability is solely specified by the matrix' internal structure and consequently it should be independent of the liquid and of the temperature. 

\subsection[Influence of confinement]{Influence of Confinement}
So far we have completely neglected that the mean pore diameters of the pore network are orders of magnitude smaller than characteristic in usual flow paths in common miniaturized fluid manipulating applications. Indeed, the pore radii are merely 10 to 100 times larger than typical molecular diameters of simple liquids like water. For that reason it is evident that some questions about the influence of the confinement on the fluid dynamics arise. In the following the two most apparent ones will be discussed.

\subsubsection{Validity of Continuum Mechanical Theory}
Up to now we have assumed the law of Hagen-Poiseuille to be valid even in pores with diameters below 10\,nm. However, one must not forget that this law is based on the principles of continuum mechanical theory, in which the behavior of a fluid is determined by collective properties such as the viscosity $\eta$ and the surface tension $\sigma$. This assumption certainly holds for ensembles of $10^{23}$ molecules. But within the pore confinement such amounts are not reached.  Assuming water molecules to be spheres with a radius of 1.5\,\AA\ in a hexagonal close-packed structure one arrives at only $1000$ molecules per cross-sectional area. As a consequence, the validity of the continuum theory has to be put into question. 

On this score especially the development of the surface force apparatus (SFA) has stimulated extensive studies over the last three decades. The mobility of water and several hydrocarbons in extremely confined films was examined by experiment \cite{Israelachvili1986, Horn1989, Raviv2001} and in theory \cite{Gupta1997}. These studies revealed a remarkable robustness of the liquids' fluidity down to nanometer and even subnanometer spatial confinement. Moreover the validity of macroscopic capillarity conceptions at the mesoscale was demonstrated \cite{Fisher1981, Fradin2000,Seveno2013, Landers2013}. The measurements presented below will provide further hints whether the concepts of viscosity still remain valid in nanopore confinement.

\subsubsection{Validity of the No-Slip Boundary Condition}  \label{subsectionNSBC}
The law of Hagen-Poiseuille implies the no-slip boundary condition. This means that the velocity of the fluid layers directly adjacent to the restricting walls equal the velocity of the walls themselves. Nowadays it is indisputable that this assumption does not hold unreservedly. Already 60 years ago Debye and Cleland introduced both slipping and sticking fluid layers at the pore walls in order to interpret their seminal experiment on liquid flow across porous \V \cite{Debye1959}. In that way, they were able to quantitatively account both for increased as well as for decreased measured flow rates (compared to the predicted ones) within their examinations of the flow of hydrocarbons through porous \Vend.

\begin{figure}[!ht]
\centering
\includegraphics*[width=.9\linewidth]{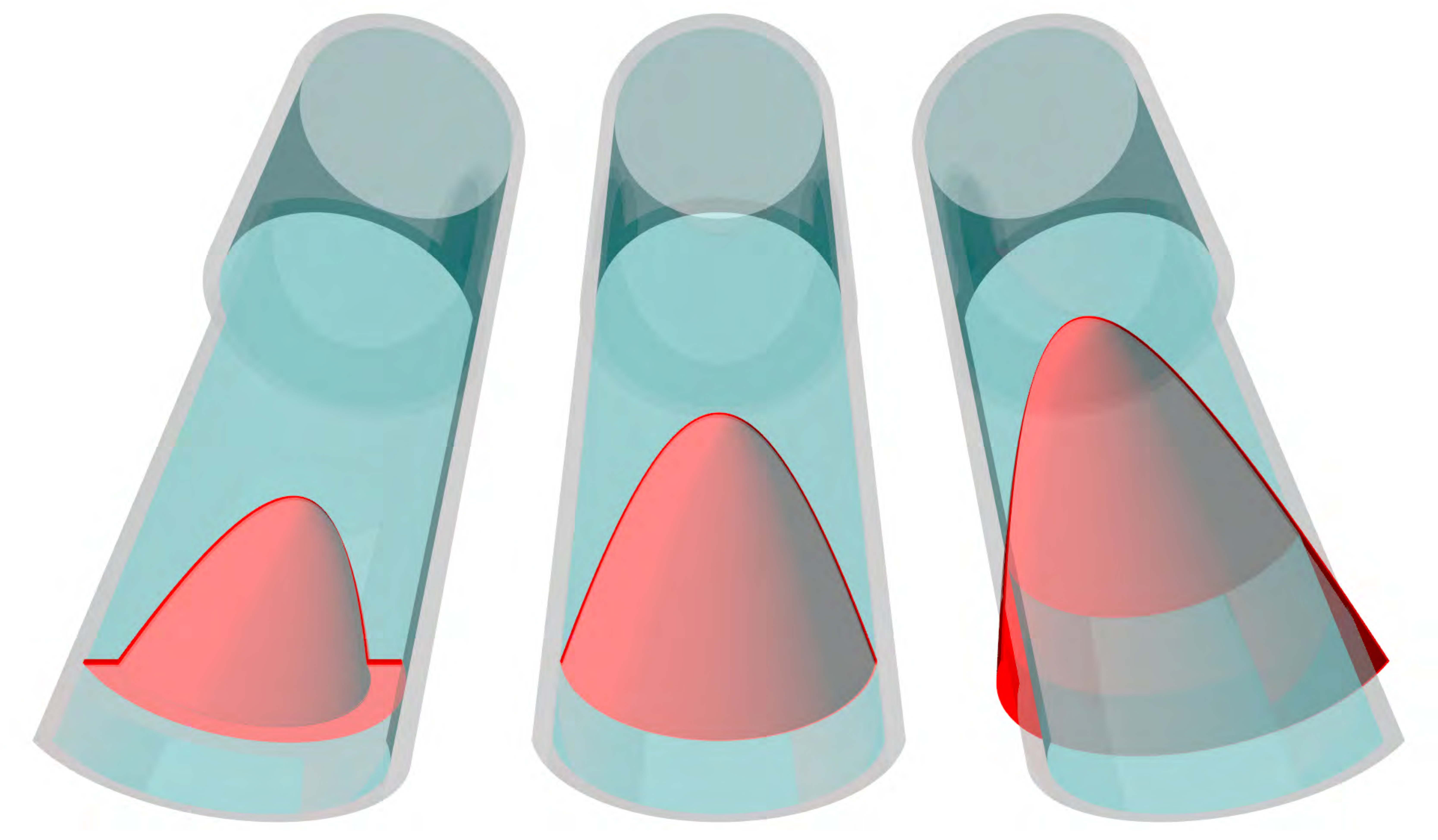}
\caption[Boundary conditions]{(Color online) Illustration of the possible boundary conditions along with the corresponding parabolic velocity profiles in a cylindrical tube with radius $\subs{r}{0}$. Mass transport takes place only where the streaming velocity is different from zero. ({left}): The reduction of the net flow rate is due to sticking layers at the pore walls, which do not participate in the mass transport. In addition the maximum velocity in the pore center is smaller than for no-slip boundary conditions ({middle}) because of the smaller hydrodynamic pore radius $\subs{r}{h}<\subs{r}{0}$. This gives rise to a further dramatic decrease in the flow rate. ({right}): In contrast, a slipping liquid with a hydrodynamic pore radius $\subs{r}{h}>\subs{r}{0}$ causes the highest streaming velocity and consequently the highest net flow rate.}
\label{stickslip}
\end{figure} 
The concepts of a sticking and of a slipping liquid compared to the traditional no-slip boundary condition are exemplified in Fig.~\ref{stickslip} for a cylindrical tube with radius $\subs{r}{0}$. The degree of slip can be quantified by the slip length $b$\label{sliplength} with $\subs{r}{0} \equiv \subs{r}{h}-b$. The hydrodynamic pore radius $\subs{r}{h}$\label{hydrodynamicradius} measures the distance from the pore center to the radius where the streaming velocity reaches zero. In this representation the sticking layer boundary condition is indicated by a negative slip length $b$ whereas a positive slip length is typical of a slip boundary condition. The standard no-slip condition yields $b=0$ meaning $\subs{r}{0} = \subs{r}{h}$. 

Because of the modified boundary conditions one has to substitute $\subs{r}{h}$ for $r$ in Eq.~\eqref{eq:HagenPoiseuille}. This procedure yields
\begin{equation}
K= \frac{\subs{\phi}{0}}{8\,\tau} \,\frac{\subs{r}{h}^4}{\subs{r}{0}^2} = \frac{\subs{\phi}{0}}{8\,\tau} \,\frac{(\subs{r}{0}+b)^4}{\subs{r}{0}^2} 
\label{eq:permeability2}
\end{equation}
for the permeability of the membrane. Equation~\eqref{eq:permeability2} illustrates the high sensitivity of $K$ on $b$, provided $b$ is on the order of or even larger than $\subs{r}{0}$. Therefore, measuring the hydraulic permeability gives direct access to the slip length $b$ for a given liquid under given conditions. 

One has to keep in mind that boundary conditions and fluid properties derived from measured flow rates are subject to a central restriction: one cannot verify the predefined parabolic shape of the velocity profile in the mesoscopic flow geometry. This is because there is no direct access to the profile itself but only to flow rates, which correspond to the velocity profile integrated over the whole pore cross-sectional area. Nevertheless, molecular dynamics simulations prove the formation of parabolic flow profiles even down to channel radii of 3 molecular diameters \cite{Todd1995, Travis1997, Binder07} and, hence, justify inferences based on this major assumption.

\subsection{Forced imbibition}
In the case of forced imbibition, where an external pressure is applied to induce liquid flow in a porous medium, the dynamics of the flow through a host of thickness $d$ and cross-sectional area $A$ (that is already completely filled with the liquid) can directly be related to Darcy's law \rel{eq:Darcy} in conjunction with the permeability $K$ according to \rel{eq:permeability2}. For a pore network with mean pore radius $\subs{r}{0}$, porosity $\subs{\phi}{0}$ and tortuosity $\tau$ and with the liquid's viscosity $\eta$, in terms of the volume flow rate this finally reads
\begin{equation}
\dot{V} = \underbrace{\frac{A\,\subs{\phi}{0}}{8\,d\,\eta\,\tau}\, \frac{\subs{r}{h}^4}{\subs{r}{0}^2}}_{\subs{C}{V}}\,\Delta p
\label{eq:MFAflow}
\end{equation}
with $\Delta p$ denoting the (externally generated) pressure drop that is applied along $d$. By determining the prefactor $\subs{C}{V}$ through a measurement of $\dot{V} (\Delta p)$ the hydrodynamic pore radius $\subs{r}{h}$ is easily accessible. 

\section{Experimental}
\subsection{Materials}
The spatial restrictions in the nanometer range were provided by the sponge-like topology of porous \V glass (Corning, code 7930). \V is virtually pure fused silica glass permeated by a three-dimensional network of interconnected, elongated pores \cite{Levitz91, Mitropoulos95, Gelb98, Huber1999}. The experiments were performed with two types of \V significantly differing in the mean pore radius $\overline{r}_0$ only, whereas they both coincide in the volume porosity $\phi_0 \approx 0.3$. The aspect ratio $a$=pore diameter/ pore length of \V glasses is between 5 and 7 \cite{Levitz91, Mitropoulos95, Gelb98}. For convenience the two types will be termed V5 ($\overline{r}_0=3.4$~nm) and V10 ($\overline{r}_0=5.0$~nm) in the following.  The matrix properties have been determined by means of nitrogen sorption isotherms performed at 77~K. 

Prior to using, we subjected them to a cleaning procedure with hydrogen peroxide and nitric acid followed by rinsing in deionized Millipore water and drying at 200~\degr C in vacuum for two days. This treatment ensures the removing of any organic contamination on the large internal surface of the samples. Until usage the samples were stored in a desiccator. 

The \V membranes are highly hydrophilic. This is a consequence of glass being a high-energy surface with chemical binding energies on the order of 1\,eV. Nearly any liquid spreads on such surfaces. This behavior can be comprehended considering the Young-Dupr\'e equation (with the indices {\textit{S}}olid, {\textit{L}}iquid and {\textit{V}}apor of the interfacial tension $\gamma$\label{interfacialtension} and the static contact angle $\subs{\theta}{0}$\label{staticcontactangle})
\begin{equation}
\subs{\gamma}{SV} = \subs{\gamma}{SL}+    \subs{\gamma}{LV} \,\cos\subs{\theta}{0} \, .
\label{eq:YoungDupre}
\end{equation}
The empirical Zisman criterion predicts that any liquid fulfilling $\subs{\gamma}{LV} < \subs{\gamma}{C}$ (with the critical surface tension $\subs{\gamma}{C}$ of the surface)\label{criticaltension} totally wets this surface. For glass it is $\subs{\gamma}{C} \approx 150\,\frac{\rm mN}{\rm m}$ \cite{Gennes2004}. Hence, even highly polarizable liquids like water spread on silica surfaces (meaning $\subs{\theta}{0} =0^\circ$). 

What is more, silica substrates provide the simple opportunity to alter the surface chemistry and thereby reduce the surface energy. This can be done by silanization \cite{Gennes2004}. Prior to silanization the samples were flushed with tri\-chloro\-methane (\ce{CHCl3}) several times. In the subsequent step they were exposed to a 1:9 mixture of dimethyldichlorosilane (\ce{Si(CH3)2Cl2}) and trichloromethane for about two hours. In presence of dimethyldichlorosilane low-energy methyl (\ce{CH3}) groups were substituted for the polar and consequently high-energy hydroxyl (\ce{OH}) groups at the glass surface. Afterwards the samples were again flushed with trichloromethane and methanol several times. 

It is important to perform this last step thoroughly since any remainder of di\-methyl\-di\-chloro\-silane in the sample potentially reverses the silanization reaction in the presence of water, e.g., from the humidity in the laboratory. In order to further minimize the risk for such a reversal reaction the samples were dried over a stream of dry nitrogen. 

The samples were characterized again by means of nitrogen sorption isotherms. They reveal a reduction in the mean pore radius of approximately 4\,\AA\, which is consistent with the thickness of the attached methyl groups at the pore walls \cite{Gruener2010}. The porosity is likewise reduced. The values are listed in \tab{tab:Vycor}. We will denote the silanized samples sV5 and sV10, respectively.

\begin{table}[!ht]
\centering
\setlength\extrarowheight{2pt}
\caption[Properties of the silanized \V batches]{Properties of the silanized \V batch as extracted from isotherm measurements.} 
\begin{tabular}{|c|c|c|}
\hline
sample batch &  mean pore radius $\subs{r}{0}$ &  volume porosity $\subs{\phi}{0}$  \\
\hline
 sV5 & $(3.0\pm 0.1)$\,nm &  $0.235\pm 0.02$  \\
\hline
\end{tabular}
\label{tab:Vycor}
\end{table}

For the permeability experiments we employed deionized Millipore water and n-hexane with a purity of 99\% as delivered from Merck.

\subsection{Hydraulic Permeability Apparatus}

\begin{figure}[!b]
\centering
\includegraphics*[width=.9\linewidth]{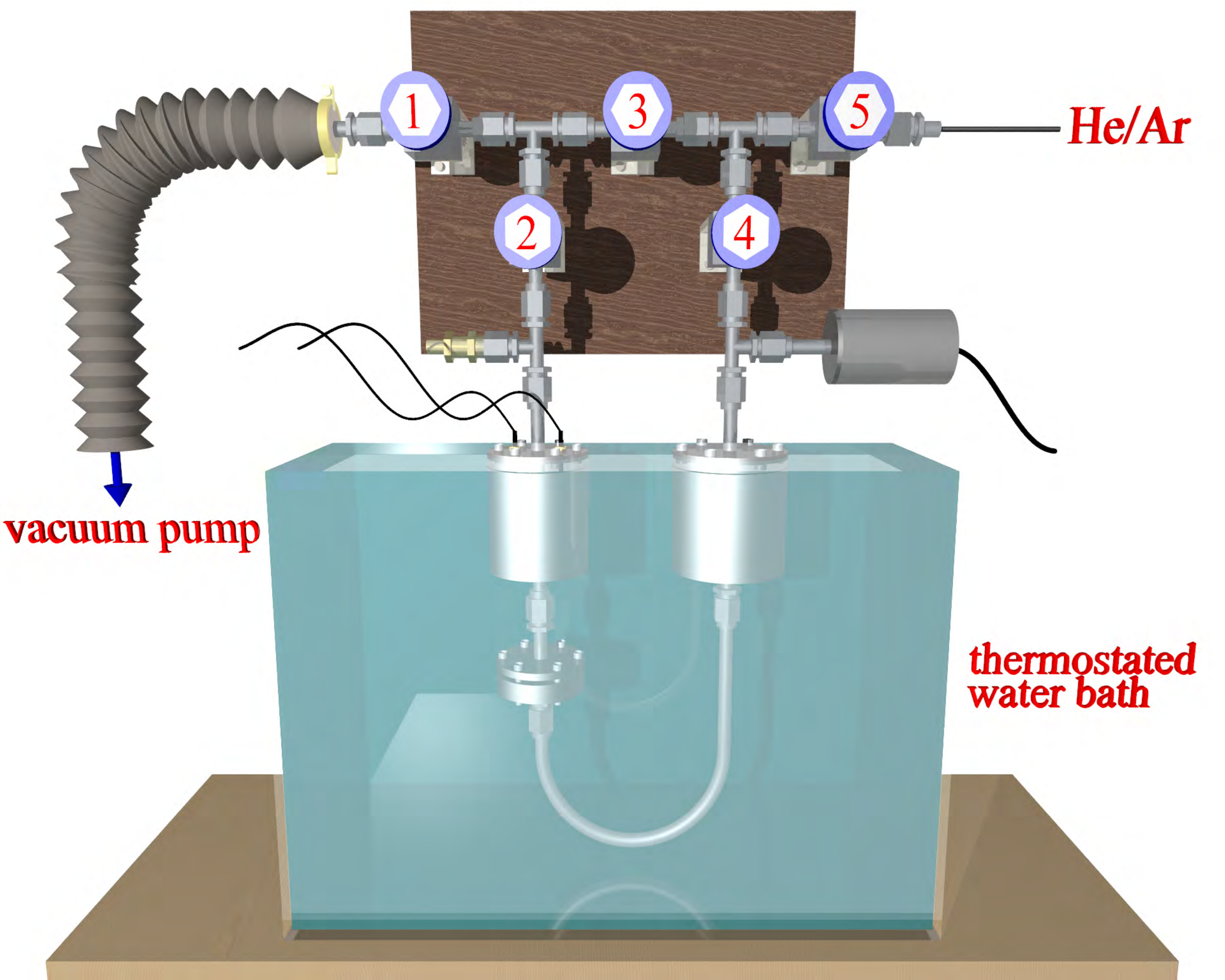}
\caption[Membrane flow apparatus (HPA)]{(Color online) Raytracing illustration of the hydraulic permeability apparatus (HPA) consisting of a gas handling and the actual flow system. The latter is temperature-controlled in a water bath. The gas handling can be evacuated by a vacuum pump and can be filled with helium or argon gas.}
\label{MFA_system}
\end{figure}
The experimental setup for the forced throughput measurements, the hydraulic permeability apparatus (HPA), is illustrated in \bild{MFA_system}. All parts liquid containing are immersed in a water bath, which can be heated up to 80\dC. The external pressure is provided by a highly pressurized gas. For this purpose the flow system is connected to a gas handling, which supplies the gas via valve 5 and 4. The valves 1, 2, and 3 permit an initial evacuation of the handling; during the measurements they are normally closed thereby separating the (right) high pressure from the (left) low pressure side. The complete setup is manufactured inhouse and made of stainless steel. This allows for maximum pressures of up to 70\,bar, which can be measured with a capacitive pressure transducer. The pressure beyond the capacitor is fixed to the upper limit of 1\,bar by means of a blow off valve.

For most measurements the liquid was pressurized with high purity helium gas (6.0). This choice was made in order to lessen the impact of a major flaw in the measuring method: the liquid stands in direct contact with the highly pressurized gas. Some imaginable consequences will be discussed in the next section. However, with the usage of an inert gas at least chemical reactions can be prevented. What is more, helium is the gas that is, at room temperature, the least soluble in water \cite{Potter1978}. In order to study a possible influence of the solubility of the gas on the dynamics argon (purity 5.7) was used as well. 

Via the supply channel the pressurized liquid in the reservoir reaches the cell with the cylindrical sample of typically $d=4$\,mm thickness and a diameter of 6\,mm. The latter is thoroughly glued into a copper sample holder using the two component, thermally conductive epoxy encapsulant Stycast 2850 with the catalyst 24LV from Emerson \& Cuming. With this procedure one must not only accomplish the task of fixing the sample, but also that of sealing the sample's side facets in order to guarantee the flow through the top and bottom facets only. Or equivalently: the procedure should ensure that the pressure drop is applied along the complete sample thickness $d$.

Beyond the sample cell the cylindrical capacitor is attached. Due to the liquid flow through the sample the liquid level in the capacitor rises thereby changing the capacitance. The latter can accurately be ascertained employing a multi frequency LCR meter (HP 4275A) at the frequency $f=500$\,kHz. This value was chosen with regard to water's high dielectric loss within the microwave range (roughly between 1\,GHz and 1\,THz), which would entail additional inaccuracies due to the strong $f$ dependency of the permittivity. For $f=500$\,kHz the dielectric constant only shows the persistent dependence on the temperature $T$.

For a direct relation between the shift in the capacitance $C$\label{capacitance} and the related change in the liquid volume $V$ in the capacitor the latter was calibrated. For this purpose its capacitance was measured while it was stepwise filled with specific amounts of the respective liquid. Since the permittivity is a function of the temperature this procedure was performed for all relevant $T$. In general, each calibration was repeated at least five times. Some of the resultant $C(V)$ curves are exemplarily shown in \bild{MFA_calibration}. 
\begin{figure}[!t]
\centering
\includegraphics*[width=.8\linewidth]{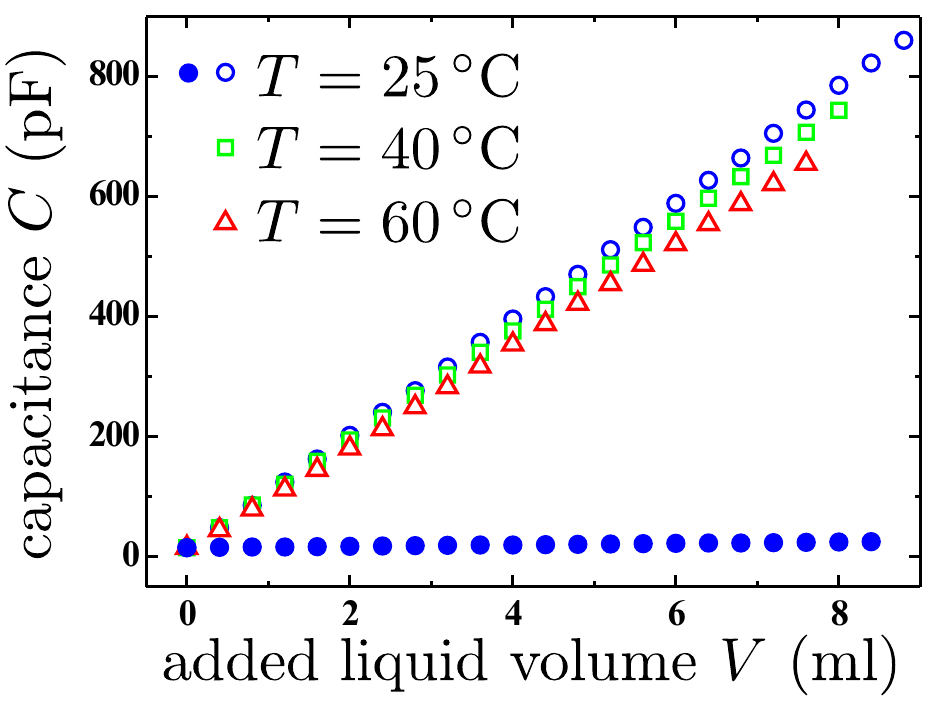}
\caption[Calibration curves of the cylindrical capacitor for water and n-hexane]{(Color online) Calibration measurements of the cylindrical capacitor for water (open symbols) and n-hexane (filled symbols) at selected temperatures. The capacitance $C$ was measured as a function of the liquid amount $V$ filled into the capacitor. For the empty capacitor it is $C\approx 15$\,pF.}
\label{MFA_calibration}
\end{figure}

The plots confirm the before-mentioned good applicability of water because of its high dielectric constant as compared with n-hexane. Additionally, the influence of the temperature is clearly recognizable: with increasing $T$ the polarizability decreases due to the enhanced microscopic mobility of the molecules. Macroscopically this behavior is expressed in terms of a decreasing permittivity of the liquid. 

One is now able to connect a certain change in $C$ with an equivalent change in $V$ via a calibration factor $\subs{C}{cal}$ that is the slope of the shown calibration curves: $\frac{{\rm d}C}{{\rm d}V}\equiv \subs{C}{cal}$\label{calibrationfactor} \cite{Gruener2010}. The flow of n-hexane was measured at 50\dC\ instead of 60\dC\ (as for water) because of the increasing noise in the proximity of its boiling point at 69\dC.

Using \rel{eq:MFAflow} this finally results in a relationship between the measured variation of the capacitance $C$ as a function of the time $t$ (at a given applied pressure gradient $\Delta p$) and the flow dynamics in confinement
\begin{equation}
\dot{C} = \subs{C}{cal} \,\dot{V} = \subs{C}{cal} \,\subs{C}{V} \, \Delta p 
\label{eq:MFAflow2}
\end{equation}
expressed in terms of the prefactor $\subs{C}{V}$ (see \rel{eq:MFAflow}). The most accurate way to deduce $\subs{C}{V}$ is extracting the slope of a $\dot{V} (\Delta p) = \frac{\dot{C}(\Delta p)}{\subs{C}{cal}} $ plot.

\section{Results and Discussion}

\subsection[Porous \V with native silica walls]{Hydraulic Transport Across Hydrophilic \Vend~with Native Silica Surfaces}
\begin{figure}[!b]
\centering
\includegraphics*[width=.8\linewidth]{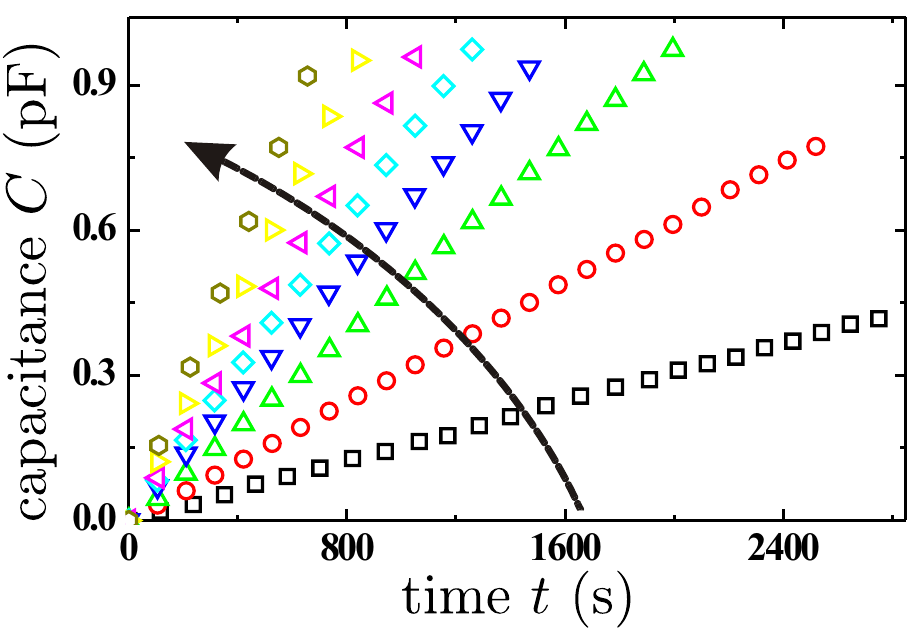}
\caption[Variation in the capacitance of the cylinder capacitor due to the flow of water through V10 for a series of applied external pressures]{(Color online) Time dependent variation in the capacitance $C$ of the cylinder capacitor due to the flow of water through V10 at 25\dC\ for a series of applied external pressures. The arrow indicates the direction of increasing $\Delta p$. The shown measurements correspond to: 8\,bar, 16\,bar, 24\,bar, 31\,bar, 37\,bar, 45\,bar, 54\,bar, and 70\,bar. The data density is reduced by a factor of 20.}
\label{MFA_flow_water}
\end{figure}
In this first part we will present results obtained from forced throughput measurements on untreated \Vend. The raw data signal of the capacitance change $C$ as a function of the time $t$ is exemplarily shown in \bild{MFA_flow_water} for the flow of water in V10 at $T=25$\dC\ and for selected applied pressures generated with helium gas. It is evident that with increasing $\Delta p$ the variation in $C$ with $t$, that is the slope $\dot C$, increases gradually. This result can directly be interpreted in terms of an increasing volume flow rate $\dot V = \frac{\dot C}{\subs{C}{cal}}$ with increasing pressure.

\begin{figure}[!b]
\centering
\includegraphics*[width=.8\linewidth]{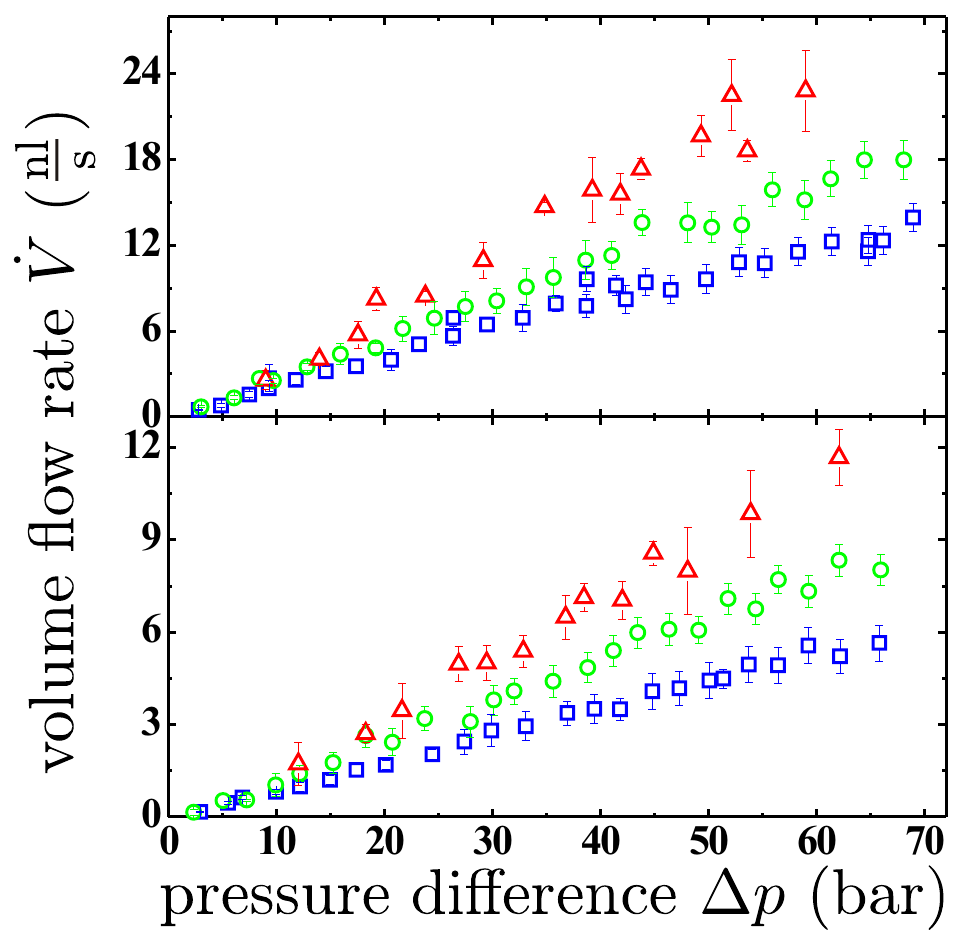}
\caption[Volume flow rates of water in porous \V as a function of the applied pressure difference]{(Color online) Volume flow rates $\dot V$ of water in V10 (upper panel) and V5 (lower panel) as a function of the applied external pressure difference $\Delta p$ at three different temperatures: 25\dC\ (square), 40\dC\ (circle), and 60\dC\ (triangle).}
\label{MFA_volumeflow_water}
\end{figure}

In \bild{MFA_volumeflow_water} some of the resultant volume flow rates $\dot V$ of water in both V5 and V10 at three different temperatures are plotted as a function of the applied external pressure $\Delta p$. The same was done for the flow of n-hexane. Some of the corresponding results are shown in \bild{MFA_volumeflow_hexane}. However, due to the rather low calibration factor of n-hexane as compared with that of water, the measuring time had to be increased in order to gain a proper signal with sufficient resolution. For that reason the overall data density is markedly reduced for n-hexane.

\begin{figure}[!t]
\centering
\includegraphics*[width=.8\linewidth]{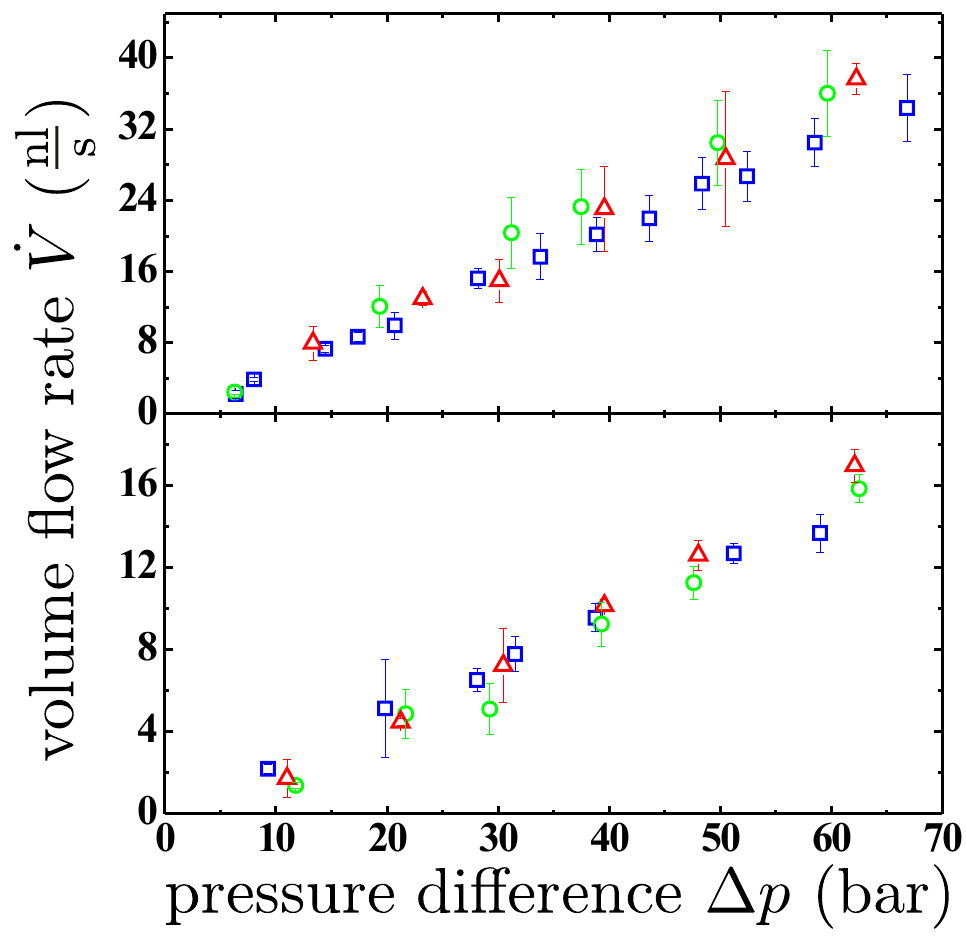}
\caption[Volume flow rates of n-hexane in porous \V as a function of the applied pressure difference]{(Color online) Volume flow rates $\dot V$ of n-hexane in V10 (upper panel) and V5 (lower panel) as a function of the applied external pressure difference $\Delta p$ at three different temperatures: 25\dC\ (squares), 40\dC\ (circle), and 50\dC\ (triangle).}
\label{MFA_volumeflow_hexane}
\end{figure}
In principle all data sets show a linear relation compliant with \rel{eq:MFAflow2}. The comparison between different temperatures implies -- at least for water -- a distinct $T$ dependence of the proportionality constant $\subs{C}{V}$: the latter increases with increasing temperature. According to \rel{eq:MFAflow} this behavior is solely determined by the temperature dependence of the liquid's viscosity. Qualitatively this is true: at higher temperatures the lower viscosities cause an increase in $\subs{C}{V}$. But, in the temperature region of interest the $T$ dependency of the viscosity of water is more distinctive than that of n-hexane. This behavior renders the effect more pronounced for water.

In a subsequent step the values of the hydrodynamic pore radii $\subs{r}{h}$ were calculated from the extracted slopes $\subs{C}{V}$. Based on the matrix properties stated earlier and on the known sample dimensions $A$ and $d$ one arrives at the slip lengths $b=\subs{r}{h}-\subs{r}{0}$ listed in \tab{tab:MFA_rh}. The error margins in $b$ represent standard deviations. Some of the volume flow rate data do not extrapolate to zero for zero pressure difference. This may result from an underestimation of the error margins and/or systematic errors in the tiny flow rates at small pressure differences, e.g. by small gas leaks in the setup or temperature drifts in the pressure gauges. Along with the error bars in the $\subs{C}{V}$ data it results in the comparably large error margins in $b$. Since the slip lengths should principally be independent of the respective liquid and the measuring temperature they allow for a more quantitative analysis and comparability of the results.
\begin{table}[!t]
\centering
\setlength\extrarowheight{2pt}
\caption[Slip lengths of water and n-hexane flowing through porous \V as extracted from forced throughput measurements]{Slip lengths $b$ (in \AA) of water and n-hexane flowing through V5 and V10, respectively, as extracted from forced throughput measurements at three different temperatures $T$ (in \dC). The liquids were pressurized with helium.} 

\begin{tabular}{|c|c|c|c|c|} 
\hline
\multirow{2}{*}{$T$}     & \multicolumn{2}{|c|}{V5}                & \multicolumn{2}{|c|}{V10}       \\  
                         & water                &  n-hexane       & water           &  n-hexane   \\ \hline
25                       &  $-2.6\pm 1.6$       & $-3.8\pm 1.8$   &  $-0.1\pm 2.2$  & $-2.8\pm 2.6$ \\
\hline
40                       &  $-1.8\pm 2.1$       & $-4.5\pm 2.2$   &  $-0.3\pm 2.9$  & $-1.9\pm 2.9$  \\
\hline
50                       &                      & $-3.5\pm 2.4$   &                 & $-3.2\pm 2.8$  \\
\hline
60                       &  $-1.3\pm 2.3$       &  				      	&  $0.7\pm 3.6$  & \\
\hline
\end{tabular}
\label{tab:MFA_rh}
\end{table}

First of all, nearly all extracted slip lengths are negative suggesting a sticking layer boundary condition in agreement with previous spontaneous imbibition experiments on water and n-alkanes \cite{Gruener2009, Gruener2009a, Gruener2015}. This indicates a compartmentation of the imbibed liquid into (1) an interfacial layer whose dynamics are mainly determined by the interaction between liquid and substrate and (2) an inner (away from the interface) region that shows the classical behavior as predicted from collective liquid properties like viscosity or surface tension.
This conclusion is supported by Molecular Dynamics studies on the glassy structure of water boundary layers in \V \cite{Gallo2000, Bonnaud2010} or more generally on layering and increased viscosities at silica \cite{Argyris2008, Xu2009a, Xu2009b} and hydrophilic surfaces \cite{Sendner2009,Bonthuis2012, Vo2015, Bhadauria2015}, by structural studies documenting a partitioning of water in a core and a surface water contribution in silica pores \cite{Bellissent-Funel2003, Erko2011} and by tip-surface measurements in purified water \cite{Li2007}. It is also consistent with beam-bending experiments on water permeability of \V \cite{Scherer2000, Vichit-Vadakan2000a} as well as pecularities in the measured thermal expansion and diffusivities of aquaous solutions confined in \V \cite{Xu2009a, Xu2009b}.

This immobile shell in the case of hexane is in agreement with the pioneering experiments on forced imbibition of n-alkanes by Debye and Cleland \cite{Debye1959}, mentioned above, as well as experimental and theoretical studies regarding the thinning of n-alkane films in the surface force apparatus \cite{Chan1985, Christenson1982, Stevens1997, Georges1993, Heinbuch1989}. 

Moreover, X-ray reflectivity studies indicate one strongly adsorbed, flat lying monolayer of hydrocarbons on silica \cite{Basu2007, Mo2005, Bai2007, Corrales2014}. Quasi-elastic neutron scattering measurements, which are sensitive to the center-of-mass self-diffusion of the n-alkanes in the pores and thus the liquid's viscosity, also indicate a partitioning of the diffusion dynamics of the molecules in the pores in two species: One component with a bulk-like self-diffusion dynamics and a second one which is immobile, sticky on the time scale probed in the neutron scattering experiment \cite{Baumert2002, Kusmin2010, Kusmin2010a, Hofmann2012}. 

However, by means of gravimetrical capillary rise measurements \cite{Gruener2009, Gruener2009a} and beam-bending experiments \cite{Scherer2000, Vichit-Vadakan2000a, Xu2009a} a thickness of the sticky layer of approximately 5\,\AA\ and 6\,\AA\, respectively, for water in \V were inferred. But the values stated in \tab{tab:MFA_rh} all deviate towards lower values and eventually $b$ turns even positive for water in V10. For water there seems also to be a marginal increase in $b$ with $T$, whereas there is no systematic dependency for n-hexane. Contrasting the results for water with the results for n-hexane it turns out that the slip lengths for water are always higher than those for the alkane. Furthermore, the values for V10 are systematically increased as compared with V5.

The bottom line of these results is that the forced imbibition dynamics are generally increased as compared to previously reported spontaneous imbibition experiments \cite{Gruener2009, Gruener2009a, Gruener2010, Gruener2011,Gruener2015}. Additionally, there are configurations regarding the flowing liquid and the substrate that seemingly facilitate higher slip lengths. This observation can be condensed as follows:
\begin{eqnarray}
 b({\rm hexane}) &<& b({\rm water})   \nonumber \\
  b({\rm V5}) &<& b({\rm V10}) \; . \nonumber
\label{eq:MFA_slip}
\end{eqnarray}
The increase of $b$ with increasing temperature for water is only vague but should not remain unmentioned at this point.

In the forced imbibition measurements the liquid stands in direct contact with the highly pressurized gas. Consequently, it is unavoidable that gas is dissolved in the liquid and thereby possibly influences the flow experiments. It has often been reported that dissolved gas modulates slip \cite{Granick03, Dammer06}. For Newtonian fluids enhanced dynamics were found to be consistent with a two-layer-fluid model, in which a layer $<1$\,nm thick, but with viscosity 10 - 20 times less than the bulk fluid, adjoins each solid surface \cite{Zhu01}. A potential mechanism to explain the genesis of this layer was discussed by Vinogradova \cite{Vinogradova99} and formalized by de Gennes \cite{Gennes2002}, who hypothesized that shear may induce nucleation of vapor bubbles; once the nucleation barrier is exceeded the bubbles grow to cover the surface, and the liquid flow takes place over this thin gas film rather than the solid surface itself. Hence, the segregation of gas at the near-surface region seems to facilitate some kind of low-density surface regions, but the nature of these is not understood well at this time.

\begin{table}[!b]
\centering
\setlength\extrarowheight{2pt}
\caption[Solubilities of helium and argon in water and n-hexane at 1\,bar for selected temperatures]{Solubilities $\mathcal S$ (in $\frac{\rm mmol}{\ell}$) of helium and argon in water \cite{Potter1978, Baranenko1989} and n-hexane \cite{Markham1941, Clever1957, Hesse1996} at 1\,bar for selected temperatures (in \dC).} 
\begin{tabular}{|c|c|c|c|c|}
\hline
\multirow{2}{*}{$T$}     & \multicolumn{2}{|c|}{helium}    & \multicolumn{2}{|c|}{argon}       \\ 
                         & water         &  n-hexane      & water           &  n-hexane   \\ 
\hline
20                       &  $0.36$       &                &  $1.55$         &              \\
\hline
25                       &               & $1.98$          &  $1.40$         & $19.50$        \\
\hline
40                       &  $0.32$       & $2.40$          &  $1.15$         & $18.30$        \\
\hline
60                       &  $0.27$       &  				     	&  $0.94$         & \\
\hline
\end{tabular}
\label{tab:solubilities1}
\end{table}

SFA measurements on tetradecane performed by Granick \etal impressively elucidate this theory \cite{Granick03}. The experiments showed that whereas no-slip behavior was obeyed when the tetradecane had been saturated with carbon dioxide gas, massive deviations from this prediction were found when the tetradecane was saturated with argon. Argon possesses only low solubility in tetradecane what may have made it more prone to segregate at the surfaces. 

According to these results and considerations the shear rate and the solubility of the gas (hereinafter denoted as $\mathcal S$)\label{gassolubility} determine the possible influence of such segregation at a near-surface region. In the following we will assess whether a process like this can be responsible for the observed peculiarities. 

First of all, an impact of the shear rate can indeed be noticed. Since for a given applied pressure difference the maximum shear rate in a channel increases with the fifth power of the channel radius, one may conclude that gas segregation, and therefore enhanced flow dynamics are more likely in V10 than in V5. This behavior could explain the higher slip lengths in V10 as compared to V5. Note that in the case of water flowing through V10, our data would even be fully compatible with the assumption of $b$=0. This finding would not be compatible with the assumption that the slip length is solely determined by adsorbed molecular layers and thus by the fluid-wall interactions, since they are identical for both matrices.

\begin{table}[!t]
\centering
\setlength\extrarowheight{2pt}
\caption[Slip lengths of water and n-hexane flowing through porous \V pressurized by different gases]{Slip lengths $b$ (in \AA) of water and n-hexane flowing through V10 pressurized by two different gases, namely helium (He) and argon (Ar).} 
\begin{tabular}{|c|c|c|c|} 
\hline
         liquid          & temperature          &  He             & Ar                 \\ 
\hline
water                    &  25\dC               & $-0.1\pm 2.2$   &  $-1.2\pm 3.3$    \\ 
\hline
\multirow{2}{*}{n-hexane}&  25\dC               & $-2.8\pm 2.6$   &  $-2.3\pm 3.1$    \\
                         &  40\dC               & $-1.9\pm 2.9$ 	&  $-1.6\pm 3.5$  \\
\hline
\end{tabular}
\label{tab:MFA_pressing_gas}
\end{table}

A potential effect caused by the gas' solubility in the respective liquid can be assessed considering the solubilities listed in \tab{tab:solubilities1}. It is obvious that for a given temperature the solubility of helium is higher in n-hexane than in water. According to the before-mentioned segregation of gas and the enhanced flow dynamics should be more likely for water than for n-hexane. This prediction coincides with the observed systematically higher slip lengths for water. Even the vague increase in $b$ with the temperature $T$ is consistent with the slight decrease in $\mathcal S$ with increasing temperature. Granick's conjecture is in accord with the observed behavior.

For an additional test some forced imbibition experiments in V10 were also carried out with argon instead of helium. According to \tab{tab:solubilities1} its solubility in water is about 4 times higher than that of helium; in n-hexane it is even up to 10 times higher. Accordingly, for both experiments one would expect smaller slip lengths as compared to the measurements with helium. For water there is indeed a slight decrease in $b$ - see \tab{tab:MFA_pressing_gas}. Though, for n-hexane the result is rather ambiguous. 

Note, however, that the experiments with argon rule out the possibility that the dissolved gas results in an "apparent" $b$ reduction mediated by a decrease in viscosity of the liquid in the pore center by the gas. Upon changing from helium to argon a pronounced viscosity drop and thus increase in $b$ would be expected in this case because of the 10-fold higher solubility of argon in the liquid. However, by contrast rather a $b$-decrease is found, in agreement with the smaller tendency for gas separation at the solid-liquid interface.

\begin{table}[!b]
\centering
\setlength\extrarowheight{2pt}
\caption[Solubilities of helium in water at 25\dC\ for selected pressures]{Solubilities $\mathcal S$ (in $\frac{\rm mmol}{\ell}$) of helium in water at 25\dC\ for selected pressures $p$ (in bar) \cite{Baranenko1989}.} 
\begin{tabular}{|c|c|c|c|c|c|c|c|c|} \hline
$ p  $        & 1    &  3   &    5   &   7.5  & 10  &    25  &    50    &  75  \\
 \hline
$\mathcal S$  & 0.36 & 1.09 &  1.77  &  2.7  & 3.53 &   8.78 &  17.5    & 26.3 \\\hline
\end{tabular}
\label{tab:solubilities2}
\end{table}

Of course, important information could be gained from experiments where the fluid is completely separated from the gas, e.g. by a flexible but gas-impermeable membrane. Despite the fact that we tried a variety of flexible membranes with varying chemical constitution, we always noticed a final gas permeability. Still, for the future it would be interesting to compare experiments with and without such membranes. A sizeable reduction of the gas dissolution could in particular be possible for argon, since for the large argon atom (in comparison to helium) flexible membranes with smaller gas permeability are available, e.g. based on polyvinylidene chloride, \cite{Liesl2002}. 
However, already the experiments by Debye and Cleland \cite{Debye1959} are in this respect an important reference for the present study, since they performed forced imbibition studies on native \V glass without the application of pressurized gases. In agreement with the spontaneous imbibition experiments discussed above, they found a smaller slip length for n-hexane in V10 than in our experiments performed under inert gas pressure, i.e. $b= -5\AA$. Hence, also a comparison with their experiments corroborates our considerations with regard to a gas-dissolution induced increase in hydraulic permeability of \V glass.

\subsection[Porous \V with silanized silica walls]{Hydraulic Transport Across Hydrophobic \Vend~with Silanized Silica Surfaces}
As outlined in the introduction the high significance of the liquid-substrate interaction in restricted geometries has been pointed out several times so far. In particular the boundary conditions are markedly influenced by the wettability of the substrate \cite{Barrat99, Pit00, Cieplak01, Tretheway02, Cho04, Schmatko05, Fetzer2007, Voronov08, Maali08}. This encourages measurements on the flow dynamics through porous \V with a modified surface chemistry. 

\begin{table}[!b]
\centering
\setlength\extrarowheight{2pt}
\caption[Slip lengths of water and n-hexane flowing through treated and untreated \Vend]{Slip lengths $b$ (in \AA) of water and n-hexane flowing through untreated and surface silanized \Vend, respectively. The liquids were pressurized with helium.} 
\begin{tabular}{|c|c|c|c|c|} 
\hline
system                            &      temperature    &  untreated         & silanized        \\ \hline
water in (s)V5                   & 25\dC               & $-2.6\pm 1.6$      &     n/a          \\  
\hline
\multirow{3}{*}{hexane in (s)V5}& 25\dC               & $-3.8\pm 1.8$      &  $0.3\pm 2.7$    \\ 
                                  & 40\dC               & $-4.5\pm 2.2$      &  $0.4\pm 2.8$    \\
                                  & 50\dC               & $-3.5\pm 2.4$ 	    &  $0.4\pm 3.8$  \\
\hline
\end{tabular}
\label{tab:MFA_silanized}
\end{table}
The results from the measurements on silanized \V compared to the values from the respective untreated sample are shown in \tab{tab:MFA_silanized} in terms of slip lengths. The value for water in sV5 is not available since even for the highest pressures applied (70\,bar) no flow through the sample could be detected. Contrary to this collapse in the dynamics of water the flow of n-hexane seems to have been even enhanced.

It is obvious that the modified surface chemistry of the porous \V samples significantly influences the dynamics of both liquids. Some basic discoveries are in high accordance with an NMR study of water and several alcohols in similarly treated \V glass \cite{Hirama96}. The inability of water to penetrate the sV5 sample must be traced back to the modified wettability of the substrate. Spontaneous imbibition could be observed for neither sV5 nor sV10. In consequence, a capillary depression caused by a contact angle $\subs{\theta}{0} > 90^\circ$ is substituted for the capillary rise mechanism. One can estimate a lower bound for $\subs{\theta}{0}$ from the finding that even pressures up to 70\,bar cannot overcome the counteracting Laplace pressure:
\begin{equation}
\cos \subs{\theta}{0} < - \frac{\Delta p \;\subs{r}{0}}{2\,\sigma} \; ,
\label{eq:dewetting_angle}
\end{equation}
thus it is $\subs{\theta}{0} > 98^\circ$. Depending on the actual methyl density of the silanized surface, water can have contact angles up to $120^\circ$ corresponding to a Laplace pressure of $\sim 240$\,bar. Therefore, the complete blocking of water penetration of the sV5 sample is not surprising at all \cite{Cruz-Chu2006}. It is rather a preeminent elucidation of the magnitude of surface forces.

The results on the flow of n-hexane in sV5 (the same sample that was used in the water experiment) can be explained by the  reduction of the surface energy of \V due to silanization. It weakens the attractive interaction between the surface and the alkane. This is expressed by the distinct disappearance of the sticking layer in favor of a classical no-slip boundary condition although, according to the Zisman criterion, the liquid should still totally wet the surface.

\section{Conclusions}

To sum up, we performed experiments on the pressure-driven flow of water and n-hexane across monolithic nanoporous \V. The hydraulic flow rates can be rationalized, if one assumes a negative velocity slip length, i.e. a sticking molecular layer. The thickness of this layer is thinner than inferred from spontaneous imbibition experiments. This observation is traced to an increased slippage at the pore wall resulting from the partial dissolution of the gases used to drive the flows.
Moreover, we verified that the wettability of the substrate deeply influences the flow dynamics and boundary conditions. The observed effects range from increasing slip lengths for n-hexane to complete blocking of the flow for water. Especially for the alkane we observe that by silanization of the pore wall, we can achieve a vanishing of the sticking boundary layer.

For the future, the study of longer alkanes, which possess higher surface tensions, would permit more detailed examinations of the influence of the wettability. Also experiments on silanized samples with larger pore diameter than examined here are planned, since they should allow us to overcome the pressure barrier for water flow in the pores and thus to study water transport in hydrophobic pores. Furthermore, the surface coating with fluorinated groups (instead of methyl groups) causes reductions of the critical surface tension down to $\sim 6\,\frac{\rm mN}{\rm m}$ \cite{Gennes2004}. By these means the interplay of surface wettability (solid-liquid interactions) and confinement and its impact on flow at the continuum limit could be further explored \cite{Vincent2015}.

\begin{acknowledgments}
This work has been supported by the German Science Foundation (DFG) within the priority program 1164, ''Nano- \& Microfluidics'' (Grant. No. Hu 850/2) and within the Collaborative Research Initiative SFB 986, ''Tailor-Made Multi-Scale Materials Systems", research area B, Hamburg.
\end{acknowledgments}
%

\end{document}